\documentstyle[12pt]{article}

\newcommand{\be}{\begin{equation}}
\newcommand{\ee}{\end{equation}}
\newcommand{\bea}{\begin{eqnarray}}
\newcommand{\eea}{\end{eqnarray}}
\newcommand{\beaa}{\begin{eqnarray*}}
\newcommand{\eeaa}{\end{eqnarray*}}
\newcommand{\ben}{\begin{enumerate}}
\newcommand{\een}{\end{enumerate}}
\newcommand{\ei}{\end{itemize}}
\newcommand{\half}{\frac{1}{2}}

\def\zzz{\mathop{\rm Z\kern -0.25em Z}\nolimits}
\def\th{\theta}

\def\bt{\beta}
\def\al{\alpha}

\def\Lm{\Lambda}
\def\lm{\lambda}
\def\eps{\epsilon}

\def\gm{\gamma}
\def\dl{\delta}
\def\Dl{\Delta}

\def\Gm{\Gamma}
\def\del{\nabla}

\def\ra{\rightarrow}
\def\half{\frac{1}{2}}
\def\reva{\frac{1}{4}}  
\def\t1{\theta_1}
\def\bar{\overline}
\def\ints{\int_0^{2\pi}}

\def\bhh{\hbar}

\begin{document}

\begin{center}
This paper has been submitted to the SIAM Journal of Applied Mathematics 
under the title \\ 
{\bf Bifurcation Analysis for Phase Transitions in Superconducting Rings \\
 with Nonuniform Thickness}\\

{\bf 1. INTRODUCTION}
\end{center}

Superconductivity is a macroscopic manifestation of quantum mechanics. This 
is noticeable in Josephson junctions and in the Little-Parks effect
\cite{lipa}, where the behavior of a ring is not determined by the local 
magnetic field, but rather by the magnetic flux which it encloses.
In the original Little-Parks experiment, the superconducting sample was a 
long cylindrical shell, intended to have uniform thickness, embedded in 
a magnetic field parallel to the axis. The sample `does not like' to 
enclose a non-integer number of magnetic flux quanta. As a consequence, 
when the flux deviates from an integer number, the transition to
superconductivity occurs at a lower temperature.  

Another interesting feature of superconductivity is its globallity. 
More precisely, a uniform material at uniform temperature will be
 superconducting at some places 
and normal at others if this nonuniform pattern minimizes the overall
 thermodynamic potential. When parts of the sample are 
normal, while the rest is superconducting, the material is said to
be in a mixed state.
A known case of mixed state is the ``intermediate state" \cite{abr,tin} for 
type~I superconductors. In this case normal and superconducting domains of 
finite volume coexist. Another known case is that of ``vortices"
 \cite{abr,tin} for type~II superconductors.
 For each vortex, the material is normal 
(i.e. the order parameter vanishes) along a line.
Using physical arguments and numerical minimizations, we recently predicted 
\cite{beru} a situation in which quantization and globallity give rise to a 
qualitatively different kind of mixed state. In this case the sample is 
doubly connected, i.e. homotopic to an annulus, 
but it is energetically advantageous to break the 
superconductivity at a suitable layer, so that the superconducting part is 
singly connected. When this is the situation, we shall say that the sample 
is in the ``singly-connected phase".

In the present paper we analyze the different kinds of functions that 
minimize the thermodynamic potential in the Little-Parks experiment. They 
depend upon the temperature and the magnetic flux, with
the deviation from uniform thickness as an additional parameter. 
The deviation from uniformity will be usually taken as a perturbation. A 
complementary analysis is provided by the
special case of shells whose thickness is piecewise constant.
We shall see that the ``phase diagram" (the boundaries between 
qualitatively different minimizers in the temperature-flux plane) contains 
several bifurcation lines and critical points. Even an infinitesimal 
deviation from uniform thickness turns out to produce a qualitative change 
in the phase diagram. We provide an analytic treatment of the minimization 
problem near the bifurcation lines and the critical points.  

In the next section we formulate our energy functional, define nondimensional
quantities and derive the Euler-Lagrange equations. In Section 3 we consider
the case of uniform rings and gain some insight into the phase transition
picture. Bifurcation lines from the normal state are studied in Section 4.
We show that there actually exist two types of bifurcations: one in which the
sample becomes superconducting everywhere, and another which involves a
mixed (singly-connected) state. The mixed state itself may bifurcate into
doubly connected superconducting states. The
location of such bifurcations in the parameters plane (magnetic flux and
temperature) is found in Section~5. 

We show that there exist two critical points where bifurcation lines
intersect. Section 6 is devoted to a detailed analysis of the various 
phases near these points. In Section 7 we summarize our results
and discuss them. In particular we propose several experimental setups
to verify our theoretical predictions.
   
\begin{center}
{\bf 2. FORMULATION}
\end{center}
We deal with a sample of superconducting material with the topology of a 
ring. The traditional treatment of the Little-Parks experiment may be found
in \cite{tin}. Here we develop a formalism which is suited for the case of
nonuniform thickness.
The free energy of the sample can be expressed by means of the 
Ginzburg-Landau theory \cite{abr}, \cite{cho}, \cite{dgp}, \cite{lali}. 
If we write the order parameter in the form $\psi=u e^{i\phi}$, then
the free energy density of the superconducting phase, relative to the 
normal phase, is
\be
g= -\al u^2+\half \bt u^4+( H-H_{\rm ext})^2/(8\pi)
 +[\bhh^2(\del u)^2+u^2(\bhh \del \phi - e^* A/c)^2]/(2m^*).
\label{gl1}
\ee
Here $e^*/2$ and $m^*/2$ are the electron charge and mass, 
$H=\del \times A$ is the magnetic field, 
$ H_{\rm ext}$ is the magnetic field which would be present in the absence
of superconductivity, $\bhh$ is Planck's constant divided by $2\pi$, $c$ is
the speed of light, and $\al,\bt$ are two material parameters. We denote
by $T_c$ the critical temperature in the sample without magnetic field and 
are interested in the region where the temperature $T$ is lower than and 
close to $T_c$. 
The parameter $\al$ is proportional to $1-T/T_c$.

To fix ideas, we shall consider a sample in the form of a cylindrical shell 
and use cylindrical coordinates $(z,r,\th)$, with the axis of the 
coordinates at the axis of the shell. We denote by $R$ the radius of the 
cylinder (measured up to the middle of the shell) and denote by $D(\th)$ 
the area obtained by cutting the shell with a plane of given $\th$.
 $D(\th)$ is the product of the ``height" (parallel to the axis) and the
``thickness" (along the radial direction). The height will be taken as 
uniform. In the original Little-Parks experiment, the height was much 
larger than the thickness; in recent experiments, the height is smaller 
than the thickness.
 There are two lengthscales \cite{dgp} in superconductivity: the magnetic
field varies significantly over a lengthscale called the {\em penetration
depth}, given by $\lm_L=(c/e^*)(m^*\bt/4\pi\al)^{1/2}$; the lengthscale 
over which the order parameter $\psi$ varies significantly is the {\em 
coherence length}  $\xi=(\bhh^2/2m^*\al)^{1/2}$. Since
$\al \rightarrow 0$ for $T  \rightarrow T_c$, both lengthscales are large 
near $T_c$. We shall see that the interesting temperatures are those where 
$\xi$ is comparable with $R$. We shall take thickness and height
much smaller than $R$, so that $\psi$ becomes a function of $\th$ only.
 (Also in the original Little-Parks 
experiment $\psi$ is independent of $z$, due to translational symmetry.)

Since $H-H_{\rm ext}$ is proportional to the current along the ring, its 
contribution to the free energy is proportional to the square of 
$D(\th)$, whereas the contribution of the other terms is linear
in $D(\th)$. Therefore, the term with $H-H_{\rm ext}$ becomes negligible
at the interesting temperatures, and will be dropped for simplicity. The
contribution of this term will be evaluated in Appendix A and, in 
particular, we shall have a criterion for the validity of this 
approximation. 

In order to write the free energy in a more convenient form, we observe
that the supercurrent $I$ along the ring is given by
\be
I=(e^*/m^*)u^2(\bhh \del \phi - e^*A/c)D(\th),
\label{scur}
\ee
and it is independent of $\th$. 
Dividing (\ref{scur}) by $u^2D$, and integrating along the ring, we obtain
an alternative representation for the supercurrent,
\be
I=2 \pi k \bhh \al e^*/(m^* \bt \Lm).
\label{scur2}
\ee
Here 
\be
 k=\frac{1}{2 \pi}[\phi(2 \pi) -\phi(0)] - \Phi /\Phi_0,
\label{flux}
\ee
 $\Phi$ is the magnetix flux, $\Phi_0=2 \pi \bhh c/e^*$ is the flux quantum,
\be
\Lm=R\int_0^{2 \pi} y^{-2} D^{-1} d \th,
\label{lm}
\ee 
and we have scaled the order parameter $y=(\bt/\al)^{1/2}u$.

We can now use the equality between both expressions for $I$ to get rid of 
the term with $\del \phi$ in (\ref{gl1}) and
write the energy functional in the form  
\be
\tilde{G}=\int_{\rm sample} \frac{\bt g }{\al^2 R}=
\int_0^{2\pi}[-y^2+\half y^4+(\frac{\xi}{R})^2 
(y^{'})^2]
D(\th) d \th+\frac{(2 \pi k \xi)^2}{R \Lm}.
\label{gl2}
\ee

Note that in obtaining (\ref{gl2}) we have not actually 
required any particular shape
of the ring. We shall keep this notation with circular appearence for
aesthetic reasons, but all our treatment remains valid for an arbitrary
loop if we define $R$ as its length divided by $2\pi$ and $\th$ as the
arc length divided by $R$. Also for aesthetic reasons, we define
$\lm=(R/\xi)^2$, multiply $\tilde{G}$ by the constant 
$\lm/\ints D(\th) d \th$,
take $R$ as the unit of length, and arrive at our model for
the free energy
\be
G[y,\phi]=\left( \int_0^{2\pi} (-\lm y^2+\half \lm y^4 +(y^{'})^2 ) 
D d \th +(2\pi k)^2\Lm^{-1} \right)/\ints D d \th.
\label{gl3}
\ee

A rigorous derivation of (\ref{gl3}) as a limit of the Ginzburg-Landau 
functional in thin rings will be published elsewhere.

While the dependence of $G$ on the amplitude $y$ is quite complex, 
$G$ depends on the phase $\phi$ only through $k^2$. For a given flux $\Phi$,
the phase $\phi$ will adjust itself to minimize $k^2$ in (\ref{flux}).

In order to minimize $G$, the order parameter has to choose between two
distinct strategies, each of which gives rise to a constraint.
\begin{itemize}
\item {\em Doubly-connected phase} (DC): The entire sample is superconducting 
($y >0$ for all $\th$) and the topology of the superconducting region is the
same as that of the sample. The constraint is that 
$[\phi(2 \pi) -\phi(0)]/(2\pi)$ must be integer and this condition restricts
the minimum value which $k^2$ can assume for a given $\Phi$. For any $\Phi$,
$k^2$ can be in the interval $[0,\frac{1}{4}]$, and therefore it will be in 
this interval when $G$ is minimized. The interval $k^2 \le \reva$ will be 
dubbed {\em the relevant region}.
However, there is much to be learned by allowing values
$k^2>\frac{1}{4}$ in our model functional $G$, and we shall study this 
situation, too. The last term in the numerator of
(\ref{gl3}) may be regarded as the energy
price that superconductivity has to pay for staying doubly-connected and
enclosing a noninteger flux.
\item {\em Singly-connected phase} (SC): In this case $y$ vanishes for some
$\th$, so that the superconducting region is singly-connected and does not
enclose the magnetic flux. $\phi(2\pi)$ and $\phi(0)$ are now unrelated, so
that their difference can adjust itself to yield $k=0$ in (\ref{flux}) and
the last term in the numerator of 
(\ref{gl3}) vanishes. We can also regard this vanishing
as due to the divergence of $\Lm$ (and then the value of $k$ is actually
irrelevant). The constraint is now that $y$ vanishes at some $\th$, and
the energy price is in the first and third term in the integrand in 
(\ref{gl3}).
We shall often consider the case in which $D$ is an increasing function of 
$\th$ for $0<\th<\pi$ and is symmetric about $\th=0$ and $\th=\pi$. We 
shall denote such $D$'s as {\em functions of type S}.
 In this case, the place with the lowest energy price for $y$ to vanish will be
at $\th=0$.
\ei

Our goal is now to find the function $y$ that minimizes $G$ for given
values of the parameters $\lm$, which is controlled by the temperature, and
$k$, which is governed by the magnetic flux. In particular, we want to find
the values of $\lm$ and $k$ for which $y(0)$ vanishes, and how $y(0)$
 approaches 0 as
these parameters vary. Note that the minimum value of $k^2$ which is 
consistent with (\ref{flux}) is a periodic function of the flux. This implies
that all physical properties will be periodic functions of the flux. This is
a fair idealisation of the experimentally known results.

Minimizing (\ref{gl3}) over all periodic functions, we obtain the
 Euler-Lagrange (EL) equation
\be
(Dy^{'})^{'}+\lm D(y-y^3)-(\frac{2\pi k}{\Lm})^2 \frac{1}{D}y^{-3}=0.
\label{ely}
\ee
The task of solving Eq.~(\ref{ely}) is complicated by the presence of the
nonlocal quantity $\Lm$. This reflects the nonlocal nature of the effect
we are discussing. For the SC phase the EL equation reduces to
\be
(Dy^{'})^{'}+\lm D(y-y^3)=0,\;\;\;y(0)=y(2\pi)=0.
\label{el0}
\ee

It is sometimes more convenient to work with $w=y^2$. 
 We therefore write the functional
$G$ and the associated EL equation in terms of $w$
\be
G[w]=\left(\int_0^{2\pi} (-\lm w+\half \lm w^2 +\frac{(w^{'})^2}{4w} ) 
D d \th +\frac{(2\pi k)^2}{\Lm}\right)/\ints Dd\th,
\label{glw}
\ee
\be
Dww^{''}+D^{'}ww^{'}-\frac{D}{2}(w^{'})^2+2\lm Dw^2(1-w)-\frac{2}{D}
(\frac{2\pi k}{\Lm})^2 =0,
\label{elw}
\ee
and now $\Lm= \int_0^{2 \pi} (wD)^{-1} d \th.$ If $w$ vanishes at some 
point, then, as before, the terms involving $\Lm$ are not present.
Adding to the numerator of (\ref{glw})
 half the integral of (\ref{elw}) divided by $w$, we
obtain that, if the EL equation is fulfilled, then the free energy is
\be
	G_{\rm EL}=-\half \lm \ints Dw^2 d\th/\ints Dd\th \ .
\label{GEL}
\ee
This shows that any nontrivial solution of the EL equation has lower free
energy than the normal state.

Since $G$ is invariant under
multiplication of $D(\th)$ by a constant, the average of $D$ can be set as
1 without loss of generality. Except for Subsection 4.3, we shall adopt 
this convention in the following.

\begin{center}
{\bf 3. UNIFORM $D$ AND OUTLINE OF GENERAL $D$}
\end{center}

The case $D$=constant is the traditional Little-Parks problem. In this case
it is not difficult to guess the solutions of (\ref{ely}) and (\ref{el0}).
For the doubly-connected phase we try the ansatz $y$=constant, which gives
\be
  w=y^2=1-k^2/\lm ,
\label{cony}
\ee
for which $G$ becomes
\be
  G_{\rm const}=-\half (\lm-k^2)^2/\lm.
\label{conG}
\ee
For the SC phase
\be
   y=\sqrt{\frac{2m}{1+m}}{\rm sn}(\sqrt{\frac{\lm}{1+m}}\th|m),
\label{SCy}
\ee
where $m$ obeys
\be
   K(m)=\pi \sqrt{\frac{\lm}{1+m}};
\label{SCK}
\ee
this gives the free energy
\be
 G_{\rm SC}=\frac{2\lm}{3(1+m)}(-\frac{2+m}{1+m}+\frac{2E(m)}{K(m)}).
\label{SCG}
\ee
In the equations above, sn is a Jacobian elliptic function and $K$ and $E$
are complete elliptic integrals \cite{abi}.

For the constant phase, we note that $y \equiv 0$ for $k^2=\lm$ and becomes 
meaningless for $k^2>\lm$. This defines a bifurcation line from the normal
to the constant phase, denoted by $\Gamma_{\rm I}$ in Fig.~1. Likewise, the SC
solution vanishes for $\lm=\reva$ and is meaningless for $\lm<\reva$.
This defines the bifurcation line $\Gamma_{\rm II}$  from the normal to the 
SC phase.
For $k=0$, $G_{\rm const}<G_{\rm SC}$, but $G_{\rm const}$ increases with 
$k^2$. $G_{\rm const}=G_{\rm SC}$ for
\be
   k^2=\reva+(1-\sqrt{2/3})(\lm-\reva)-(\lm-\reva)^2/\sqrt{486}
      +5(\lm-\reva)^3/54^{3/2}+...
\label{firstorder}
\ee
Except for the point ${\rm P}=(k^2=\reva,\lm=\reva)$,
 where $G_{\rm const}=G_{\rm SC}=0$, equation (\ref{firstorder}) describes a
line in the region $k^2>\reva$. Therefore,
the global minimizer in the relevant region is the constant phase.

Let us check now whether there are bifurcations from the constant phase. For
this purpose, let us anticipate the nonuniform case and write
\be
	D=1+\eps D_1(\th)
\label{epsd}
\ee
where $\eps$ is a small positive number and the deviation from uniformity is
defined such that $\ints D_1(\th)d\th=0$. Expanding the order parameter in
 the form
\be
w=1-k^2/\lm+\eps w_1+..., 
\label{AC}
\ee
with $\lm>k^2$, Eq.~(\ref{elw}) gives to first order in $\eps$ 
a linear nonlocal differential equation for $w_1$. The nonlocal term
vanishes by periodicity, and we obtain
\be
w''_1+(6k^2-2\lm)w_1=-4k^2(1-k^2/\lm)D_1.
\label{GammaIV}
\ee

We note now that, even in the uniform case $D_1 \equiv 0$, (\ref{GammaIV})
has nontrivial periodic solutions for $6k^2-2\lm=n^2$, with $n$ an integer.
Nontrivial solutions for $n=0$ are incompatible with the periodicity of the
nonlocal equation and $6k^2-2\lm=1$ defines the line $\Gamma_{\rm IV}$, where
the constant solution bifurcates to solutions of EL that have a sinusoidal
perturbation superimposed on the constant background. 
The cases $n>1$ define lines which are far from 
the relevant region, and will not be considered.

There is still another bifurcation line which we know of. We shall see in
Section 5 that at the line $k^2=\reva$ there is a bifurcation from the SC
phase. We denote this line by $\Gamma_{\rm III}$. 

We have verified numerically that in the region between $\Gamma_{\rm III}$ and
$\Gamma_{\rm IV}$ there exists a solution of EL in addition to (\ref{cony})
and (\ref{SCy}), which approaches continuously the SC and the constant
solution at the respective bifurcation lines. This bridging solution is never
a minimizer of $G$. The numerical procedure is described in Appendix~B.

The point P at $\lm=k^2=\reva$ is a multiple point that sits at the 
intersection of four
bifurcation lines. As a consequence, we expect EL to be highly degenerate at
P. To get a better insight into the degeneracy, recall that
near P one can assume that $w$ is very small. Therefore we temporarily drop the
 cubic term in (\ref{elw}) (cf. (\ref{elwc}) below). The modified equation  
has a three-parameter solution at P (for constant 
$D$):
\be
	w_P=A[1-\gm {\rm cos}(\th-\th_0)].
\label{wP}
\ee
The arbitrary phase $\th_0$ is due to the independence of $D$ on $\th$, the
arbitrary normalization $A$ is due to the bifurcation at $\Gamma_{\rm I}$ or
$\Gamma_{\rm II}$, and $0 \le \gm \le 1$ is due to the bifurcation at 
$\Gamma_{\rm III}$ and $\Gamma_{\rm IV}$. Since $\Gamma_{\rm III}$ and 
$\Gamma_{\rm IV}$
cut each other, this family of solutions bridges between SC ($\gm=1$) 
and the constant solution ($\gm=0$) at a single point.

Our range of ``interesting temperatures" focuses on the vicinity of P, i.e. 
where the coherence length is of the order of the diameter of the ring.

Let us now outline the case of nonuniform $D$. If $D$ and $w$ can be 
expressed by (\ref{epsd}) and (\ref{AC}), then (\ref{GammaIV}) can be 
solved by Fourier analysis. For conciseness, we take
\be
  D_1=\sum_{n=1}^{\infty} \bt_n \cos n \th.
\label{FourD}
\ee
Addition of odd terms is straightforward. We make the important observation
that if $D$ is of type $S$ then $D_1$ has the form (\ref{FourD}), with 
$\bt_1 <0$. If $D$ is not of type $S$, we still define $\bt_1=\frac{1}{\pi}
\ints D_1 \cos\th d\th$ and take $\bt_1 <0$. Unless the first harmonic of 
$D_1$ vanishes, this choice of sign is just a choice of the origin of
$\th$.

The solution of (\ref{GammaIV})
is 
\be
w_1=4k^2(1-\frac{k^2}{\lm}) \sum_{n=1}^{\infty} 
		\frac{\bt_n \cos n \th}{n^2-6k^2+2\lm} \ .
\label{FourAC}
\ee

This expansion provides an approximation for the order parameter in most of 
the relevant region, but
 is not valid near some of the bifurcation lines. We shall see that,
due to the highly
degenerate character of P, even an infinitesimal deviation from constant $D$
has a profound influence on the phase diagram. We have cumulative evidence
that, for nonuniform thickness, P splits and the phase diagram looks as in
Fig.~2. In particular, there are regions where the doubly-connected phase
resembles more the SC than the constant solution. The following sections are
devoted to the justification of Fig.~2 to the study of $w(\th)$ near the
bifurcations.

We mention that in the phase diagrams (Figs. 1 and 2) we have marked only 
the two global minimizers we know of, singly and doubly connected. We have 
also met other qualitatively distinct solutions of EL, such as the 
``bridging solution" mentioned above, solutions with several oscillations, 
and solutions with a minimum where $D$ is largest. As far as we have
tested, these solutions are never global minimizers, and we do not report 
on them in detail.

An experimentally interesting quantity is the supercurrent.
Far from the bifurcations we introduce
(\ref{epsd}) and (\ref{AC}) into (\ref{scur2}) and note that $D_1$ and $w_1$ 
average to zero. We obtain
\be
 I=k\bhh\al e^*(\lm-k^2)\bar{D}/(m^*\bt\lm)+O(\eps^2),
\label{scur3}
\ee
where $\bar{D}$, the average of $D$, is now required. For fixed $\lm$,
 this current attains 
its maximum value at $k=\sqrt{\lm /3}$. Using the definition of $\lm$, this 
maximum is, up to $O(\eps^2)$,
\be
 I_{\rm max}=\left(\frac{2\al}{3}\right)^{3/2}\frac{e^*\bar{D}}{\bt
 \sqrt{m^*}} \ .
\label{Imax}
\ee

\begin{center}
{\bf 4. BIFURCATIONS FROM THE NORMAL PHASE (N)}
\end{center}

When $\lm=0$ and $k \neq 0$, the global minimizer is obviously $y \equiv 0$. 
Increasing $\lm$, with fixed $k$, we expect to reach a critical value
$\lm_c$, corresponding to the critical temperature, at which the
normal solution will bifurcate to a minimizer $y_c$ which does not vanish
identically. Here the subscript $c$ stands for either I or II.

In order to perform the bifurcation analysis of (\ref{elw}), we introduce
a small parameter $\eta$, and expand 
\be
\lm=\lm_c+\eta\lm_{\eta 1}...,\;\;w=\eta w_c+\eta^2 w_{\eta 2}+...
\label{expand}
\ee
We also write an expansion for $\Lm$
\be
\Lm^{-2}=\frac{\eta^2}{\Lm_c^2}+\frac{2\eta^3}{\Lm_c^3}\ints\frac{
w_{\eta 2}d\th}{Dw_c^2}+...,
\label{explm}
\ee
with
$\Lm_c \equiv  \Lm(w_c)=\int_0^{2\pi}(Dw_c)^{-1}d \th$.
We thus obtain a sequence of equations for $w_c, w_{\eta 2},....$ 
The leading order term $w_c$ obeys
\be
Dw_cw_c^{''}+D^{'}w_cw_c^{'}-\frac{D}{2}(w_c^{'})^2+2\lm_c Dw_c^2-
\frac{2}{D}(\frac{2\pi k}{\Lm_c})^2 =0 \ .
\label{elwc}
\ee
The second order term $w_{\eta2}$ satisfies
\be
L_{\eta}(w_{\eta 2})=2 \lm_cDw_c^3-2 \lm_{\eta 1} D w_c^2,
\label{elw2}
\ee
where the linear operator $L_{\eta}$ is defined by
\be
L_{\eta}(u)=Dw_cu''+(D'w_c-Dw'_c)u'+((Dw'_c)^{'}+4\lm_cDw_c)u-
\frac{4(2\pi k)^2}{D\Lm_c^3}\ints\frac{ud\th}{Dw_c^2} \ .
\label{Leta}
\ee
We remark that Equation (\ref{elwc}) can be written in the form
 $L_{\eta}(w_c)=0$.
Notice also that (\ref{elwc}) is invariant under multiplication of $w_c$ by
a constant. 

We shall now study the curve $\lm_c=\lm_c(k)$, and the corresponding
structure of $w_c$. Equation (\ref{elwc}) is difficult to analyse, as
it is nonlocal in a non-standard way. This difficulty can be overcome in 
the case of weakly nonuniform rings. We shall support our findings  
 by solving (\ref{elwc}) exactly, with arbitrarily large
nonunifomities, for the case where $D$ is piecewise constant.
The high degeneracy near P implies that two types of expansion are needed.
We shall first study the bifurcation from the normal state to the doubly
connected phase away from P. The solution in the vicinity of P is 
postponed to Section 6. The bifurcation to the singly connected phase is a 
particularly simple case.

\begin{center}
{\bf 4.1 Bifurcation from Normal to DC}
\end{center}

We are after a nontrivial solution of Equation (\ref{elwc}).
We choose the normalization such that 
\be
\ints w_c d\th=2\pi
\label{Norm}
\ee
and consider almost uniform thickness, given by (\ref{epsd}). Using our
knowledge for the case of constant $D$, we write 
$w_c=1+\eps w_1+\eps^2 w_2+...$ and $\lm_c=k^2+\eps \lm_{c1}+
\eps^2 \lm_{c2}+...$.
Expanding (\ref{elwc}) and using (\ref{Norm}), gives (to first order in 
$\eps$) 
\be
w_1^{''}+4k^2 w_1=-2\lm_{c1}-4k^2 D_1.
\label{AC1}
\ee
Integrating and applying (\ref{Norm}) and periodicity, 
we find that $\lm_{c1}=0$.
Using (\ref{FourD}), Fourier analysis of (\ref{AC1}) gives
\be
w_1=\sum_{n=1}^{\infty} \frac{\bt_n \cos n\th}{\frac{n^2}{4k^2}-1}.
\label{w1rsol}
\ee 
Except for normalization, this is the limit of (\ref{FourAC}) for 
$\lm \ra k^2$.

Expanding to second order in $\eps$ gives, after some manipulation,
\be
\lm_{c2}=\frac{k^2}{2}
      \sum_{n=1}^{\infty} \frac{\bt_n^2}{\frac{4k^2}{n^2}-1}.
\label{w2sol}
\ee 

We see that, no matter how small $\eps$ is, the approximation scheme breaks
down for $k^2=\reva$. This was expected, since at this point $\Gm_{\rm I}$
intersects other bifurcation lines.

\begin{center}
{\bf 4.2 Bifurcation from Normal to SC}
\end{center}

As in the previous subsection, $\lm_c$ is characterized by the fact that
(\ref{el0}), or (\ref{elw}) without the term involving $\Lm$, have a 
nontrivial solution, but the nonlinear term $\lm D y^3$ or $2\lm D w^3$ is
negligible.
It is easier to use the $y$ formulation. Writing 
$y_c=y_0+\eps y_1+\eps^2 y_2+...$ and 
$\lm_c=\reva+\eps \lm_{c1}+\eps^2 \lm_{c2}+...$, we obtain a sequence of 
equations
of the form
\be
   L_{\rm II}(y_0)=0;\;\;L_{\rm II}(y_i)=f_i,
\label{LSC}
\ee
where $f_i$ is expressed in terms of the functions $y_j$ with $j<i$ and
\be 
	  L_{\rm II}(y_i) \equiv y_i^{''}+\reva y_i \ .
\label{LSCdef}
\ee
$L_{\rm II}$ is linear and self adjoint in the space of functions that vanish
at 0 and $2\pi$, with inner product $(u,v)=\ints uvd\th$. Therefore, $f_i$
must fulfill Fredholm's condition \cite{fred} for solvability of the 
nonhomogeneous equation:
\be
	\ints f_i y_0 d\th=0 \ .
\label{fredII}
\ee
This determines $\lm_{ci}$. Taking $D$ of type $S$  and 
recalling the form (\ref{FourD}) and
 the normalization (\ref{Norm}), we obtain
\bea
   y_0 &=& \sqrt{2}\sin(\th/2), \nonumber \\
   \lm_{c1} &=& \reva \bt_1 , \nonumber \\
   y_1 &=& -\frac{\sqrt{2}}{4}\sum_{n=1}^{\infty} \left(
\frac{\bt_{n+1}}{n}+\frac{\bt_n}{n+1} \right)\sin((n+\half)\th), \nonumber \\
   \lm_{c2} &=& \frac{3}{32}\bt_1^2-\frac{1}{8}\sum_{n=2}^{\infty} \left[
   \frac{n^2}{n^2-1}\bt_n^2+\bt_n \bt_{n-1} \right] \ .
\label{SCresults}
\eea
Note that, since $\bt_1<0$, $\lm_c<\reva$.

\begin{center}
{\bf 4.3 Piecewise Constant Thickness - an Exact Solution}
\end{center}

We note from (\ref{w1rsol}) that the first harmonic diverges for 
$k^2 \rightarrow \reva$. Therefore, $w_1(0)$ will be a large negative
quantity and, for $k^2$ sufficiently close to $\reva$, $\eps w_1(0)$
will not be small and the expansion of Subsection 4.1 breaks down.
We conjecture that, for arbitrary nonuniformities that have a first harmonic
component, $w(0)/\ints wd\th$ becomes 0 as $k^2 \rightarrow \reva$ (along
the line $\Gm_{\rm I})$.
To support this conjecture and in order to have a test field for our
approximations, we consider a special class of nonuniform rings, such that
(\ref{elwc}) can be solved exactly. In Section~6 we shall verify this 
conjecture for the case of weak arbitrary nonuniformities.

Let $D$ have the form
\be
D=\left\{ \begin{array}{ll}
d & \;\;\th < \pi/2 \\
1 & \;\;\pi/2 < \th < \pi
\end{array} \right.
\label{piece}
\ee
with $d < 1$, and $D$ symmetric about $\th=0$ and about $\th=\pi$. 
In each interval of the ring $D$ is constant, so that differentiating
(\ref{elwc}) and dividing by $w_c$ gives (in each interval)
\be
w_c^{'''}+4\lm_cw_c'=0, 
\label{simplew}
\ee
which has general solution of the form
$w_c=A_i[1-\gm_i {\rm cos}2\sqrt{\lm_c}(\th-\th_i)]$. 
We are left with 7 parameters: $A_i$, $\gm_i$, $\th_i$ and $\lm_c$. These
are determined by normalization, by having the constant $(2\pi k/\Lm_c)^2/D$
in (\ref{elwc}) given
by (\ref{lm}) in both intervals, and by continuity of $w_c$
and $Dw_c'$ at $\th=\pi/2$ and $\th=3\pi/2$. 
Discontinuity of any of these quantities would lead to a
divergence of $w_c$ or $w_c'$ (with the discontinuity in $D$
regarded as the limit of a fast change). 

Using the symmetry of the problem 
about 0 and $\pi$, using continuity of $Dw_c'$ and the fact that 
$D^2(w_c w''_c+2\lm_cw_c^2)$ has the same value at $\th=0$ and $\th=\pi$,
 taking $w_c(0) \le w_c(\pi)$
and choosing a simple normalization, we write
\be
w_c=\left\{ \begin{array}{ll}
1-\gm \cos 2\sqrt{\lm_c} \th & \;\;\th <\pi/2 \\
d +d\gm \cos 2\sqrt{\lm_c} (\pi-\th) & \;\;\pi/2 < \th < \pi 
\end{array} \right.
\label{exactwc}
\ee
The requirement of continuity of $w_c$ is now
\be
\gm \cos(\pi \sqrt{\lm_c})=\frac{1-d}{1+d}.
\label{gme}
\ee
Substituting (\ref{exactwc}) into (\ref{lm}) and (\ref{elwc}) gives, 
after some trigonometric manipulations,
\be
\pi k=\arctan(\sqrt{\frac{1+\gm}{1-\gm}} \tan \frac{\pi \sqrt{\lm_c}}{2})
+\arctan(\sqrt{\frac{1-\gm}{1+\gm}}\tan \frac{\pi \sqrt{\lm_c}}{2}).
\label{lme}
\ee

Equations (\ref{gme}) and (\ref{lme}) determine the eigenvalue $\lm_c$ and
$w_c(0)=1-\gm$. The solution that describes the line $\Gm_{\rm I}$ is
\bea
\lm_c&=&\left[\frac{1}{\pi}\arcsin\left(\frac{2\sqrt{d}}{1+d}\sin \pi k
\right)\right]^2  \\
\label{simplmI}
\gm &=& (1-d)/\sqrt{(1+d)^2-4d\sin^2 \pi k} \ .
\label{simplgmI}
\eea

For $k \ra \half$, (\ref{simplgmI}) gives $\gm \ra 1$, 
that is, $w_c$ vanishes at $\th=0$, which is the center of the thin portion
of the ring. This confirms our conjecture.
 This situation describes the point P$_1$ in Fig.~2.
 
Writing $d=1-\eps$, assuming that $k$ is bounded away from $\half$,
 and expanding in $\eps$ gives 
$\lm_c-k^2=-(k/4\pi) \tan(\pi k) \eps^2$, $\gm=\eps/(2\cos\pi k)$, in agreement
with Subsection 4.1. However,
if we also write $k=\half-\dl$ and take the limit	 
$\eps \ra 0$ with a fixed ratio $\dl/\eps$, then $\gm=O(1)$. 
More precisely, we find
$\gm=1/\sqrt{1+(2\pi\dl/\eps)^2}$ and $\lm_c-\reva=-\eps/(2\pi\gm)$.

\begin{center}
{\bf 4.4 The Bifurcating Branch ($\lm>\lm_c$)}
\end{center}

The operator $L_{\eta}$ defined in (\ref{Leta}) is self adjoint in the
 space of periodic functions under the inner product
\be
(u,v)=\int_0^{2\pi}\frac{uv}{w_c^2}d \th.
\label{inner}
\ee
Therefore, application of Fredholm's condition to (\ref{elw2}) gives
\be
   \lm_{\eta 1}=\lm_c\ints Dw_c^2d\th / \ints Dw_cd\th .
\label{lmeta1}
\ee
This determines the value of $\eta$ in (\ref{expand}). It is easy to show 
that (\ref{lmeta1}) holds for the SC solution, too.

Substituting now (\ref{expand}) into (\ref{GEL}) we find that to leading
order the free energy is 
\be
-\frac{(\lm-\lm_c)^2(\ints Dw_cd\th)^2}{4\pi\lm_c\ints Dw_c^2d\th} .
\label{freeeta}
\ee

\begin{center}
{\bf 5. BIFURCATIONS FROM THE SINGLY CONNECTED PHASE}
\end{center}

The present section deals with an essential property of our model, and the
main results are valid for arbitrary shape of $D(\th)$, i.e., the
nonuniformity in
the thickness is not necessarily small. For simplicity, we shall assume
again that $D$ is of type $S$.

In the SC solution $w(0)=0$ and near a bifurcation from it $w(0)$ is small.
The existence and nature of the bifurcations from the SC solution stem from
the divergence of $\Lm$ as $w(0) \ra 0$. 

Let us first assume that $D(\th)$ is smooth in the region near $\th=0$.
Let us write for shorthand
\be
f(\th)=D(\th)w(\th), \;\; \tilde{f}(\th)=f(0)+\half f''(0)\th^2,
\label{f}
\ee
where we have used the symmetry of the problem to include 
only even powers of $\th$, and let us 
separate $\Lm$ into a divergent and a regular part:
\be
\half\Lm=\int_0^{\infty}\frac{d\th}{\tilde{f}}+\half\Lm_{\rm R},
\label{separate}
\ee
with
\be
\half\Lm_{\rm R}=-\int_{\pi}^{\infty}\frac{d\th}{\tilde{f}}+
\int_0^{\pi}\frac{\tilde{f}-f}{f\tilde{f}}d\th.
\label{Lamdar}
\ee
$\Lm_{\rm R}$ remains finite for $w(0) \ra 0$. Performing the integration
in (\ref{separate}), 
\be
\Lm=\frac{\pi}{\sqrt{\half f(0)f''(0)}}+\Lm_{\rm R}.
\label{laplace}
\ee

We set now $\th=0$ in (\ref{elw}), note that $w'(0)=0$, introduce 
(\ref{laplace}) and (\ref{f}), and expand to order $w(0)^{1/2}$.
This gives
\be
\sqrt{2w(0)w''(0)}D(0)\Lm_{\rm R}=\pi(1-(2k)^{-2}).
\label{190}
\ee
The bifurcation line is obtained by taking $w(0) \ra 0$. This gives 
$k=\half$, independently of $\lm$. 

If $\Lm_{\rm R}>0$ (resp. $\Lm_{\rm R}<0$), then (\ref{190}) can be satisfied
with $w(0)>0$ only for $k^2>\reva$ (resp. $k^2<\reva$). 
We can therefore expect
that for $\Lm_{\rm R}<0$ there is a bifurcation from the SC to the DC phase
and, for $\Lm_{\rm R}>0$, to the bridging solution in the irrelevant region.
Numerical solution of (\ref{elw}) for $k^2 \approx \reva$ and various values
of $\lm$ confirms this scenario. 

If the thickness is almost uniform, $w(\th)$ in the SC phase
can be approximated by means of (\ref{SCresults}) 
and (\ref{elw2}). Once $w$ is known, $\Lm_{\rm R}$ can be evaluated.
Keeping one correction term in $\eps$ and in $\eta$, we obtain
\be
\Lm_{\rm R}=\frac{\pi}{2}\left(1+\frac{3\eps\bt_1\lm_c}{\lm-\lm_c}\right).
\label{195}
\ee
Recalling that $\bt_1 <0$, we see that $\Lm_{\rm R}<0$ for $\lm$ 
sufficiently close to $\lm_c$, but
eventualy changes sign at a point which we may identify as P$_2$ in Fig.~2.
Taking $\lm_c$ from (\ref{SCresults}), we find that P$_2$ is located at
$\lm=\reva-\eps\bt_1/2$.

If the thickness is uniform, then $\Lm_{\rm R}=0$ at P and is positive for
$\lm>\lm_c$, in agreement with Fig.~1.

Let us finally discuss the case where $D'(\th)$ is discontinuous at $\th=0$,
as in the example of Ref.~\cite{beru}. Due to the continuity of $Dw'$,
the discontinuity of $f'(\th)$ will be of order $w(0)$. The same will be
true for the coefficient of $\th^3$ in the expansion of $f$. Therefore,
(\ref{f}) can still be used and (\ref{Lamdar}) does not diverge as $w(0) \ra 0$.

\begin{center}
{\bf 6. THE REGION NEAR THE CRITICAL POINTS}
\end{center}

We shall again restrict ourselves to the case where $D$ is almost uniform
(cf. (\ref{epsd})), define $\dl=\reva-k^2$ and $\lm_1=(\lm-\reva)/\eps$, and
expand around the point P:
\be
   w=\eta(w_P+\eps w_1+...).
\label{expandP}
\ee
The expansion is in powers of $\eps$ only, but we explore the region where
$\lm_1$,
$\dl/\eps$ and $\eta/\eps$ are finite and the terms in the expansion depend
on these parameters. The leading order term
$w_P$ is given by the expression (\ref{wP}), where without loss
of generality we set $A=1$ and the other parameters are still undetermined.
At this stage it will be instructive to consider $D$ which is 
not necessarily symmetric about	0 and $\pi$.

Substituting (\ref{expandP}) into (\ref{elw}) leads to the equation 
$L_P(w_1)=f_1$, with
\be
L_P(u) \equiv w_Pu^{''}-w_P^{'}u^{'}+u-\frac{\Dl^{3/2}}{2\pi}
\int_0^{2\pi}\frac{u}{w_P^2}d\th,
\label{eqw1}
\ee
\be
f_1=-\Dl D_1-w_Pw_P^{'}D_1^{'}-2\lm_1w_P^2+ \frac{\eta}{2\eps}w_p^3+
\frac{\Dl^{3/2}}{2\pi}
\int_0^{2\pi}\frac{D_1}{w_P}d\th-2\Dl \frac{\dl}{\eps},
\label{fP}
\ee
and $\Dl=1-\gm^2$.
								
As in the situations considered above, $L_P$ is linear and
self adjoint under the inner product $(u,v)=\int_0^{2\pi}(uv/w_P^2)d \th$.
Also, it is easy to check that $L_P(w_P)=0$. 
The unusual feature is that there
are other solutions as well, that is, ${\rm Ker}(L_P)$ is not
one dimensional. In fact, ${\rm Ker}(L_P)$ is three dimensional. It is spanned 
by the three
functions $\{1,w_P,w_P^{'}\}$, corresponding to the three parameters
in $w_P$.  

Therefore we obtain a set of three solvability conditions 
\be
\int_0^{2\pi}\frac{f_1}{w_P^2}d \th =
\int_0^{2\pi}\frac{f_1}{w_P}d \th =
\int_0^{2\pi}\frac{f_1w_P^{'}}{w_P^2}d \th=0.
\label{solva}
\ee
The last condition implies
\be
\ints D_1\sin(\th-\th_0)d\th=0.
\label{con3}
\ee
This equation determines the position $\th=\th_0$ at 
which $w_P$ has a minimum and may
eventually vanish. Note that $\th_0$ is determined by the geometry only, and
does not vary with the temperature or the magnetic field. After having 
obtained (\ref{con3}), we can redefine $\th$ and set $\th_0=0$ without loss of
generality.

The other two conditions in (\ref{solva}) reduce to
\be
\lm_1+ \Dl^{-1/2}\frac{\dl}{\eps}=\frac{\eta}{4\eps}
\label{con1}
\ee
and
\be
4\lm_1+4\Dl^{1/2}\frac{\dl}{\eps}-\gm \bt_1-(1+\half\gm^2)\frac{\eta}
{\eps}=0.
\label{con2}
\ee
These two equations determine $\gm$ and $\eta$. The following
formula that can be derived from (\ref{con1})-(\ref{con2}) 
will be useful later on: 
\be
\frac{\eta}{\eps}=\frac{8\lm_1}{3}-\frac{2\bt_1}{3\gm}.
\label{eta}
\ee

\begin{center}
{\bf 6.1 The Bifurcation Line $\Gm_{\rm I}$ Near the Point P$_1$}
\end{center}

This is obtained by setting $\eta=0$ in (\ref{con2}) and (\ref{con1}). The
lowest value of $\lm$ for which both equations are fulfilled is obtained for
\be
\lm_1=\lm_{c1}=-\reva\sqrt{\bt_1^2+(4\dl/\eps)^2},
\label{lm1}
\ee
and
\be
\gm=-\frac{\bt_1}{\sqrt{\bt_1^2+(4\dl/\eps)^2}}=
\frac{\bt_1}{4\lm_{c1}}.
\label{gm}
\ee
The limit considered at the end of Subsection 4.3 is a special case
of (\ref{gm}), with $\bt_1=-2/\pi$.
 It is also interesting to note that (\ref{lm1}) can be extrapolated to the 
case that $\dl$ is not small and it still reproduces to order $\eps$ the 
result obtained for $\lm_c$ in Subsection 4.1.

We note that no matter how small $\eps$ is, i.e. even for arbitrarily 
weak nonuniformities, $\gm \ra 1$ as $\dl \ra 0$. Hence we deal here
with a pattern selection mechanism, where the SC pattern is selected 
among all the infinitely many solutions (\ref{wP}). 

\begin{center}
{\bf 6.2 Evaluation of $w_1$}
\end{center}

For conciseness, we shall again consider $D$ of 
the form (\ref{FourD}). Fourier analysis
of the equation $L_P(w_1)=f_1$ and use of (\ref{eta}) yield
$$w_1=\sum_{n=0}^{\infty}a_n\cos n\th $$ with
$$a_2=\frac{\gm\bt_1}{3}+\frac{\bt_2}{3}-\frac{\gm\bt_3}{4}
-\frac{\gm^2\lm_1}{3}$$ and
\be
a_n=\frac{\gm}{2n}(\bt_{n-1}-\bt_{n+1})+\frac{\bt_n}{n^2-1}
\label{an}
\ee
for $n \ge 3$. The coefficients
$a_0$ and $a_1$ remain undetermined to the present order in the
expansion (\ref{expandP}); this is the same situation we met in 
Eq.~(\ref{wP}), where $A$ and $\gm$ remained undetermined.

\begin{center}
{\bf 6.3 The Region Near the Point P$_2$}
\end{center}

We know from Section 5 that near $(k^2=\reva, \lm=\reva -\eps\bt_1/2)$ there
is a special point.
The goal of this subsection is to obtain the behavior of 
$w(0) \approx \eta w_P(0)$ near this point. For this purpose we eliminate 
$\eta$ from (\ref{con1}) and (\ref{con2}) and substitute $\gm=1-w_P(0)$.
This gives an equation for $w_P(0)$:
\be
\frac{\dl}{\eps}=\frac{1}{3}\frac{\sqrt{2-w_P(0)}}{1-w_P(0)}w_P(0)^{1/2}
(\lm_1 w_P(0)-(\lm_1+\frac{\bt_1}{2})).
\label{wP2}
\ee
Near P$_2$, $w_P(0)<<1$, hence it is negligible compared to 1 or 2. 
However, in the relevant regime $\lm_1 w_P(0)$ is of the same
order as $\lm_1+\bt_1/2$. 
Therefore, (\ref{wP2}) becomes a third degree equation for
$w_P(0)^{1/2}$. The solution of this equation provides the missing information
about $w(\th)$ to present order in the region near P$_2$.

 We see from
(\ref{wP2}) that the relevant region $\reva-k^2=\dl \ge 0$ corresponds to
$w_P(0) \ge 1+\bt_1/(2\lm_1)$.
In order to characterize the solutions of (\ref{wP2}), we have drawn in 
Fig.~3 a contour plot of $\dl(w_P(0),\lm_1)$. At the axis $w_P(0)=0$, 
$\dl=0$. For $\lm_1<-\bt_1/2$, all the factors in (\ref{wP2}) are positive,
 $\dl$ increases monotonically with $w_P(0)$,
and there is only one $w_P(0)$ for a given $\dl$. However, for 
$\lm_1>-\bt_1/2$, there is initially a valley into the irrelevant region,
 that is, $\dl$ is negative - a situation which cannot correspond to the 
minimal energy.
The level $\dl=0$ is reached again for $w_P(0)=1+\bt_1/(2\lm_1)$, and this 
is found to be the global minimizer. This situation corresponds to the ray 
to the right of P$_2$ in Fig.~2. The line
$\Gm_{\rm IV}$ is characterized by $\frac{\partial\dl}{\partial w_P(0)}=0$; 
this 
happens at $\dl/\eps \approx -(1/(3\sqrt{\lm_1}))(2(2\lm_1+\bt_1)/3)^{3/2}$.

\begin{center}
{\bf 6.4 Higher Order Treatment}
\end{center}

The first order expansion in (\ref{expandP}) leaves the $0^{\rm th}$ and
$1^{\rm st}$ harmonic of $w(\th)$ undetermined for the doubly connected 
phase. This may not be a grave shortcoming, unless they diverge and their
contribution is actually of $O(1)$. We might especially suspect that
the behavior of the entire $w(0)$ near P$_2$ is qualitatively different from 
that of $w_P(0)$. We therefore proceed to evaluate these harmonics near 
P$_2$.

In order to keep the analysis simple, we now restrict ourselves to $D(\th)$
of the form $D=1+\eps \bt_1 \cos\th$ with ($\bt_1 < 0$). 
Cases of nonuniformities that are neither smooth nor small will be reported 
elsewhere. Using (\ref{expandP}), (\ref{eta}) and
(\ref{an}) sufficiently close to $k^2=1/4, \lm_1=-\bt_1/2$ and recalling 
that $\gm=1$ at P$_2$, we can write
\be
	w=-2\bt_1\eps(1+p\eps-\cos\th)(1+q\eps-\bt_1\eps\cos\th)+O(\eps^3),
\label{variation}
\ee
with $p=a_0+a_1+\bt_1/2$ and $q=-a_1-\bt_1$. Our strategy is now to use
(\ref{variation}) without the correction term as a variational form, and to 
find the parameters $p$ and $q$ that minimize $G$ in (\ref{glw}). Since
$w(0)$ is proportional to $p$, we are especially interested in the behavior
of $p(\lm_1,\dl)$ at this minimum.

Eq.~(\ref{glw}) contains two integrals that require careful handling. One
of them is $\int D(w')^2/w d\th$, which contains a vanishing denominator for
$p\eps \ra 0$. In this case we perform the integral without taking $p\eps$
as a small quantity, and expansion on it is performed only after the
integration. 

The second delicate integral is $\Lm$. Using the periodicity and the behavior
of cos$\th$ in the complex plane, we can write $\Lm=\int(wD)^{-1}d\th$, 
where the integral is around the rectangle with
vertices at $-\pi/2$, $3\pi/2$, $3\pi/2+\infty i$ and $-\pi/2+\infty i$.
Since $wD$ is the product of three simple factors, it is easy to find the
poles and residues of the integrand, and $\Lm$ can be evaluated exactly.

We write now $k^2=\reva-\dl$, with $\dl/\eps^{2.5}<$C$<\infty$,
$\lm=1/4-\eps\bt_1/2+\eps^2 \tilde{\lm}$, and expand $G$ up to $O(\eps^3)$.
We obtain a polynomial in $\sqrt{p}$ and $q$. Minimizing this expression
gives
\be
  8\tilde{\lm}+2\bt_1(2p+3q)-9\bt_1^2=0
\label{q}
\ee
and
\be
\frac{\dl}{\eps^{2.5}}=\frac{\sqrt{2}}{3}p^{1/2}(
-\frac{\bt_1}{2}p-\tilde{\lm}+\frac{3\bt_1^2}{8}).
\label{p}
\ee
One immediately recognizes that (\ref{p}) describes the same behavior as
(\ref{wP2}), but the point P$_2$ is shifted to
\be
	\lm_{\rm P_2}=\reva-\half\eps\bt_1+\frac{3}{8}(\eps\bt_1)^2.
\label{P2}
\ee

\begin{center}
{\bf 7. DISCUSSION}
\end{center}

We have analytically obtained the phase diagram 
 in the entire temperature-magnetic
flux plane, as schematically drawn in Fig.~2. 
Some of our results are valid for small deviations from uniform thickness, 
some results are valid for arbitrary shape of $D(\th)$ and some are valid 
for piecewise constant thickness; nevertheless, all the results are 
consistent among themselves and with Fig.~2. In addition, we have 
numerically tested the predicted dependence of the minimum value $w(0)$ of 
the order parameter as the physical parameters $\lm$ and $k$ are varied.
An integrative look of all the cases considered makes it 
reasonable to expect that the results found here will remain qualitatively
correct for almost any form of the thickness $D(\th)$.

The results obtained here enable us to check the numerical findings 
obtained in
Ref.~\cite{beru} for $k^2=\reva$ and for a particular form of $D(\th)$.
In \cite{beru} we compared the free energies of the SC and the DC phases. 
Upper bounds for these free energies were obtained by using sequences of 
Pad\'e approximants of increasing length as variational forms. $\tilde{G}$ 
in equation~(\ref{gl2}) was minimized in the space of the coefficients of 
these approximants. In \cite{beru} we did not suspect that the DC phase 
goes continuously into the SC phase at the line $\Gm_{\rm III}$. Since at 
this line both phases become the same, our previous approach was clearly 
inappropriate.
Indeed, there is a region in which the singly-connected phase is 
thermodynamically favorable. We also confirm here
 that along the line $k^2=\reva$
there are {\em two} phase transitions as the temperature is varied below
$T_c$. However, the transition at P$_2$ is not discontinuous (first order),
as concluded from our early numerical results, but rather a continuous (second
order) transition after which $w(0)$ increases at the rate $dw(0)/d\lm=4$.

In the relevant region ($k^2 \le \reva$) the stability domain of the 
singly-connected phase is restricted to a line segment, and has measure zero.
This property might have applications for calibration purposes. However,
the transition to the doubly-connected phase is continuous, and near this
segment $w(0)$ will be small.

A simple experimental verification of our predictions could be performed by
measuring the supercurrent by means of a magnetometer, as in 
Ref.~\cite{silver}. Fig.~4 shows numerical evaluations of the 
supercurrent, given by (\ref{scur2}),
as functions of the magnetic flux for three different temperatures close to
P$_2$.

Another experimental verification of our predictions could be obtained by
detecting the current fluctuations \cite{osc}. In the doubly-connected
phase, for $\lm>\lm_{{\rm P}_2}$, the supercurrent does not vanish when the
flux is a half-integer number of quanta ($k^2=\reva$). As a consequence, this
situation is degenerate, since the supercurrent can flow in either sense.
Near P$_2$, the energy barrier between both minima of $G$ is small, and
the supercurrent fluctuates. However,
for $\lm \leq \lm_{{\rm P}_2}$ the supercurrent vanishes, and fluctuations 
should disappear, while the sample remains superconducting.

 A third way to 
detect whether $G$ has one or two minima is by the absence or presence of 
hysteresis as the magnetic field is varied: for $\lm>\lm_{{\rm P}_2}$, if 
the flow $\Phi$ is slowly raised and crosses a half integer of quanta 
$\Phi_0$, then according to (\ref{flux}) and (\ref{gl3}), $k$ ought to jump 
from $\half$ to $-\half$ in order to minimize the free energy.  
Following (\ref{scur2}) and (\ref{scur}), this implies a jump in the order 
parameter. Since the order parameter stays in a local minimum of the free 
energy and the sample `does not know' that there is a still lower minimum, 
the jump in the order parameter and the inversion of $I$ may occur only 
after $k$ is larger than $\half$. If the change in $\Phi$ is now reversed, 
the order parameter will jump back at a lower value of $\Phi$, after $k$ is 
smaller than $\half$. We argue that this hysteresis phenomenon should not 
occur for $\lm \leq \lm_{{\rm P}_2}$. Although at $k=\pm\half$ there is 
still a jump in the phase $\phi$, equation~(\ref{scur}) suggests that this 
jump concentrates where $u$ vanishes, and is thus not physically 
observable.

Our analysis for almost uniform thickness shows that the important part of
the nonuniformity is its first harmonic. This might explain why the effects 
predicted here were not observed in \cite{silver} for samples with a weak
link (a small region where $D(\th)<<1$), which might be thought to be 
extremely nonuniform. When the deviation from uniformity is not small, 
Eq.~(\ref{SCresults}) suggests that coupling between different harmonics 
becomes important.

From a mathematical point of view, the the Little-Parks problem poses
a nonstandard minimization problem:
\be
{\rm Min}\;\;G[y;k]=\int_0^{2\pi} (y^{'})^2 D d \th +(2\pi k)^2\Lm^{-1}
\label{nsm}
\ee
subject to the constraint $\int_0^{2\pi} y^2 D d\th=1$. 
It might appear that this is a very specific problem, but it actually stems 
from very general principles: the Ginzburg-Landau equation (which
 is nothing but a
slowly varying small order parameter expansion), conservation of current and
single-valuedness of the order parameter. In general, for a multiply 
connected sample, we shall have a problem with a larger parameter-space.
Instead of one $k$ we shall have a $k_i$ for each hole in the sample that
can enclose a flux and we shall have a $\Lm_i$ for the supercurrent around
that hole.
\newpage
\begin{center}
{\bf ACKNOWLEDGMENT}
\end{center}

This research was supported by the US-Israel Binational Science Foundation. 
J.~B. was also supported by GIF and DFG.

\begin{center}
{\bf Appendix A. Influence of the Expelled Flux}
\end{center}

We deal now with the contribution of the term $(H-H_{\rm ext})^2$
 in (\ref{gl1}),
which has so far been neglected. Let us first consider the case where the
sample is a long circular cylinder. Then it is well known
that $H-H_{\rm ext}=4\pi I/(hc)$, where $h>>R$ is the height of the 
cylinder.
This additional magnetic field exists inside the cylinder, i.e., in a volume
 $\pi R^2h$. Therefore, its contribution to $G$ is
\be
G_H=4k^2\pi^3 \kappa^{-2} h^{-1}\lm\Lm^{-2}/\ints Dd\th ,
\label{A1}
\ee
where we have used (\ref{scur2}), $R=1$, and $\kappa=\lm_L/\xi$ is
 the Ginzburg-Landau parameter.
 Note that $G_H$ is proportional to the thickness $D/h$, and we are no longer
free to choose the normalization of $D$.

If the sample is not a long cylinder, then $G_H$ has to be divided by the
self-inductance of the cylinder and multiplied by the actual
self-inductance of the sample. For simplicity, let us stick to (\ref{A1}).
The influence of $G_H$ in the EL equation is that the nonlocal term in
(\ref{elw}) is multiplied by the factor
 $(1+2\pi\kappa^{-2}h^{-1}\lm\Lm^{-1})$. Therefore, a sufficient condition 
for this correction to be uninportant is $\kappa^{-2}\lm D/h<<R$.

We are mainly interested in the influence of $G_H$ when $\Lm$ diverges.
Following the steps of Section~5, we reach again (\ref{190}), but $\Lm_R$
has to be replaced by $\Lm_R-\pi\lm/(h \kappa^2)$. Therefore, the bifurcation
from SC is still along the line $k^2=\reva$, but the point P$_2$ moves to
higher values of $\lm$.

\begin{center}
{\bf Appendix B. Numerical Solution of Equation (\ref{elw})}
\end{center}

We used the built-in MATHEMATICA functions. The procedure is as follows.
\begin{enumerate}
\item Fix $w(0)$, $\lm$ and $C=2(2\pi k/\Lm)^2$.
\item With these values fixed, we can use NDSolve to evaluate $w(\th)$. 
This function requires the initial values of $w$. $w(0)$ is given and 
$w'(0)=0$. One has to be careful not to pick values of $C$ which are too 
large, since they lead to $w(\pi)>>1$.
\item The true value of $C$ is now obtained by requiring periodicity. If 
$w$ is expected to be of type $S$, we can use the function FindRoot to 
require, for instance, that $w(3)-w(2\pi-3)=0$. (This difference is now 
regarded as a function of $C$.)
\item After the true $C$ and the corresponding $w(\th)$ are known, we 
obtain $\Lm$ by numeric integration. We then evaluate
 $k^2=C\Lm^2/(8\pi^2)$. In this way we obtain $k^2$ as a function of $\lm$ 
and $w(0)$.
\end{enumerate}

Figure B1 shows a sequence of solutions of equation (\ref{elw}) with varying 
values of $w(0)$.

\newpage
\begin{center}
Figure Captions
\end{center}

Fig. 1: Phase diagram for a ring of uniform thickness. $\lm$ is 
proportional to $T_c-T$ and $k$ measures the deviation from an 
integer number of magnetic flux quanta. The dashed lines denote 
bifurcations between solutions of the Euler-Lagrange equation which are not 
global minimizers. The line with uneven dashing denotes first order 
(discontinuous) transition between the singly and the doubly connected 
phases.

Fig. 2: Phase diagram when the thickness is not uniform. P splits into two 
critical points, P$_1$ and P$_2$, and the portion of the the bifurcation 
line $\Gm_{\rm III}$ between these points is physically relevant.

Fig. 3: Contour plot of $\dl=\reva-k^2$. $\dl$ increases toward the lower 
right. At the upper left there is a ``valley", where $\dl<0$. The line
 $\Gm_{\rm IV}$ in Fig.~2 ``goes down" along this ``valley".

Fig. 4: Supercurrent as a function of $k$ for several values of the 
parameter $\lm=(R/\xi)^2$. For $\lm \le \lm_{{\rm P}_2}$, the supercurrent 
vanishes at $k=\half$. For $\lm=\lm_{{\rm P}_2}$, the supercurrent rises 
with infinite slope. For larger $\lm$, $I(\half) \neq 0$. $I_{\rm max}$ is 
given by Eq.~(\ref{Imax}).

Fig. B1: $w(\th)$ and $k$ as functions of $w(0)$. The value obtained for 
$k$ is shown next to the corresponding curve. For the present illustration 
we took uniform $D$ and $\lm=1$. This is the ``bridging solution", 
discussed in Section~3. The extreme situations were taken from (\ref{cony}) 
and (\ref{SCy}). If $D$ is weakly nonuniform and $k \neq 0$, then the DC 
phase looks qualitatively as this bridging solution.
\end{document}